\documentclass[twocolumn,twoside,preprintnumbers,amsmath,amssymb,showkeys]{revtex4}
\usepackage{epsfig}
\usepackage{graphicx}
\usepackage{dcolumn}
\usepackage{bm}

\usepackage{fancyhdr}

\usepackage{pslatex}

\newcommand{\bd}{\begin{equation}}
\newcommand{\ed}{\end{equation}}

\def \lta {\mathrel{\vcenter
     {\hbox{$<$}\nointerlineskip\hbox{$\sim$}}}}

\begin{document}

\title{{\Large New Solutions to the Hierarchy Problem\footnote{
Presented at the First Latin American Workshop on High Energy
Phenomenology, Porto Alegre, December 3-5 2005}}}

\author{Gustavo Burdman}

\affiliation{Instituto de F\'{i}sica, Universidade de S\~{a}o Paulo\\
Cidade Universitaria, S\~{a}o Paulo SP 05508-900, Brazil}


\received{on 2 of June, 2006}

\begin{abstract}
After summarizing the status of the Standard Model, we focus on the Hierarchy Problem 
and why we believe this strongly suggests the need for new physics at the TeV scale. 
We then concentrate on theories with extra dimensions and their possible manifestations
at this scale.


\end{abstract}

\maketitle

\thispagestyle{fancy}
\setcounter{page}{1}

\section{Introduction}
The Standard Model (SM) is a very successful description of particle physics up to the 
weak scale. It agrees with all the experimental data we have our disposal today.
However, it has several shortcomings. We do not know the origin of the fermion masses, 
which in the 
SM arise from the Yukawa interactions with the Higgs doublet. 
These are extremely varied, with 
a Yukawa coupling for the top of order one, and for the electron about $10^6$ times 
smaller. 
This constitutes the fermion mass hierarchy problem. Its solution might lie at high energy 
scales
so that experiments at the TeV scale  might not necessarily explore it. 
Another question unanswered by the SM, refers to the so called gauge hierarchy problem: 
why is the 
weak scale so much smaller than  the Planck scale. We believe that the answer to 
this problem is at
the TeV scale and therefore accessible to the Large Hadron Collider (LHC). 
We will first briefly summarize the status of the 
SM in Section~\ref{sm}. In Section~\ref{hierarchy} we will expose the gauge 
hierarchy problem
in more detail. In Section~\ref{extra} we will presents a review of  solutions to the 
gauge hierarchy problem involving compact extra spatial dimensions.

\section{The Standard Model}
\label{sm}
The SM gauge group, $SU(2)_L\times U(1)_Y$, must be spontaneously broken to 
electromagnetism. 
This is achieved  through the Higgs mechanism. The spontaneous breaking
results in a set of $3$ Nambu-Goldstone Bosons (NGB), which are not physical states 
since in the   unitary gauge they are gauged away and appear as the longitudinal components
of the $W^\pm$ and the $Z^0$. Within
the SM, a complex scalar doublet, corresponding to $4$ degrees of freedom, results 
in one of them being left out, while the other are the NGBs.  
But in general, we can think of this as analogous to superconductivity: the Higgs mechanism
results in a superconducting phase, where the order parameter of the phase is a 
scalar field,
which may or may not be an elementary particle. As a result, the weak interactions become
short range, just as  electromagnetism becomes short range inside a superconductor, 
by getting 
an effective photon mass. 

The first question is: at what scale should this phase transition  occur ? 
In order to answer this question within the SM, we can look --for instance-- 
at the scattering 
of massive gauge bosons. These amplitudes (without including the Higgs) grow like
\begin{equation}
\frac{s}{M_W^2}
\end{equation}
violate unitarity unless something happens before $1$~TeV.
In the context of the SM, this means that we need $m_h<O(1)$~TeV to restore unitarity. 

The SM introduces the scalar doublet which has a Lagrangian:
\bd
{\cal L} = (D_\mu{\Phi})^\dagger 
\,(D^\mu{\Phi}) 
- V({\Phi})
\ed
with the covariant derivative of the scalar doublet is given by
\bd
D_\mu\Phi = \left(\partial_\mu + igt^a W^a_\mu + \frac{ig'}{2} B_\mu \right)\Phi
\ed
The vacuum expectation value (VEV) $\langle\Phi^T\rangle = (0~~~~v/\sqrt{2})$ breaks
the SM gauge group down to $U(1)_{\rm EM}$, leaving the photon massless and giving 
the weak gauge bosons masses: 
\bd
M_W = \frac{g\,v}{2}; ~~~~~~~~~~~~~ M_Z = \frac{\sqrt{g^2 + g^{'2}}\, v}{2} 
\ed
The minimization of the Higgs potential results in $m_h = \sqrt{2\,\lambda} v$, with 
$\lambda$ the Higgs boson self-coupling.


\begin{figure}[htbp]
\includegraphics[width=8cm]{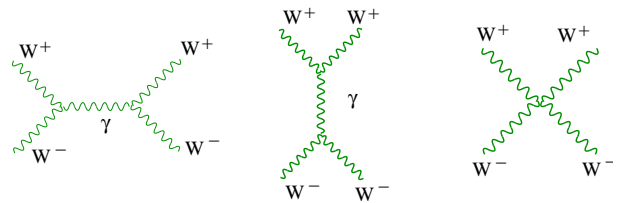}
\caption{Contributions to $WW$ scattering amplitude. It violates unitarity at about 
1~TeV}
\label{unit}
\end{figure}

\begin{figure}[htbp]
\includegraphics[width=7cm,height=10cm]{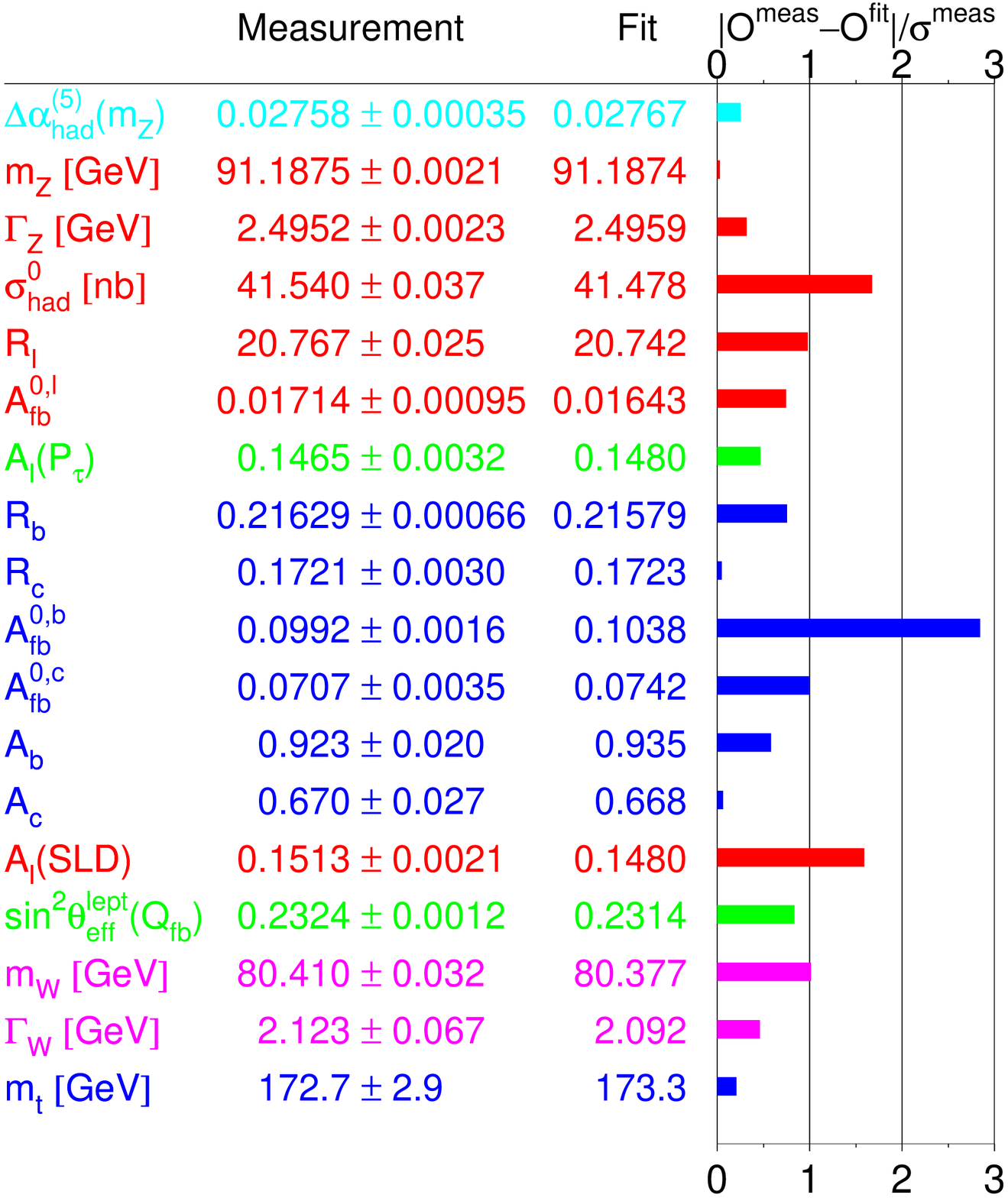}
\vskip-1.5cm
\caption{Comparison of electroweak observables with the SM fit (Summer 2005)
}
\label{pull05}
\end{figure}

\begin{figure}
\includegraphics[width=7cm,height=7cm]{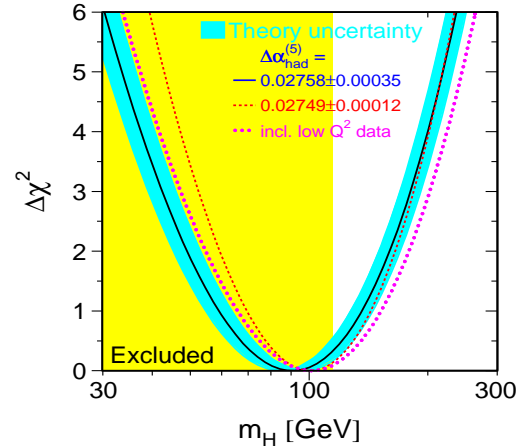}
\caption{The Higgs mass constraint in the SM. 
}
\label{chisq}
\end{figure}

The SM has been extremely successful when compared with experiment. The discovery of neutral 
currents in the 70's, and the discovery of the $W$ and $Z$ gauge bosons, where just the 
prelude to the spectacularly precise tests started in the 90's at LEP and SLAC. 
A large number of electroweak observables are predicted with the input of 
only three parameters. The fundamental parameters of the theory, $g$, $g'$ and $v$, can 
be traded by quantities that are more accurately known, typically $M_Z$, $\alpha$ and
$G_F$. 
In Figure~(\ref{pull05}) we see a sample of the observables and the agreement with the 
SM fit is very good. 
The  precision of the measurements has reach the sub-percent level, and in order
to make predictions matching this one needs to go to one loop calculations. 
This means that the measurements are sensitive to heavy physics running around loops.
The most important example is the Higgs itself. Since it has not been observed, been 
sensitive to its loop effects is very interesting. The sensitivity of the 
electroweak observables 
to the Higgs mass is only logarithmic. Yet, the precision is enough to turn 
this sensitivity 
into a bound on the Higgs mass. In Figure~(\ref{chisq}) we show the latest fit 
of the Higgs mass in the SM. The $\chi^2$ has a minimum at around $100$~GeV, although 
the direct search bound is $m_h>114$~GeV. The indirect limit implies that 
$m_h\leq 246$~GeV at $95\%$ C.L.
Thus, we conclude that the SM is in very good agreement with experiment 
and that the Higgs must be light.

Then why do we want to have physics
beyond the SM ? The SM leaves many questions unanswered, and its answers to others are not
very satisfactory. For instance, in the SM fermion masses arise as a consequence of 
Yukawa couplings to the Higgs. But since there is a huge hierarchy of masses (e.g. 
$m_e/m_t \simeq 10^{-6}$), then there should be a huge hierarchy of Yukawa couplings.  
This is the fermion mass hierarchy problem. 
The SM gauge group being a product group, plus the fact that the couplings 
get quite close to each other at high energies, suggest grand unification. 
What is the GUT group $G \subset SU(3)_c\times SU(2)_L\times U(1)_{\rm Y}$ ? 
What is the origin of the baryon asymmetry in the universe, or of dark matter 
and dark energy ?
Why is the cosmological constant so small ? So there is no shortage of shortcomings. 
However, we do not know at what energy scales these questions are answered. 
It could be that
the physics associated with them lies at a scale unreachable experimentally, 
at least for the 
moment.

\section{The Hierarchy Problem}
\label{hierarchy}
However, the question that requires physics beyond the SM at the TeV, is the so called
gauge hierarchy problem. One way to state it is simply by asking: why is the weak scale
($\simeq 100$~GeV) 
so much smaller than the Planck scale ($\sim 10^{19}$~GeV)? The weak scale is 
given by the VEV of the Higgs, $v\simeq 246$~GeV, the only dimensionfull parameter in 
the SM.  However, it is not naturally stable under radiative corrections. 
If we consider the radiative corrections to the Higgs mass, coming from its couplings to 
gauge bosons, Yukawa couplings to fermions and its self-couplings, result in a
quadratic sensitivity to the ultraviolet cutoff.
Thus, if the SM were valid up to the Planck scale, then $m_h$ and therefore the 
minimum of the Higgs, potential, $v$, would be driven to the Planck scale by the 
radiative corrections. To avoid this, one has to adjust the Higgs bare mass in 
the SM Lagrangian to one part in $10^{17}$. This is quite unnatural, and is what we call
the gauge hierarchy problem.

In order for physics beyond the SM to regulate the Higgs mass, and restore naturalness, 
its energy scale must be around the TeV. Several alternative theories are potential 
candidates to do this job. Most of them, imply that new physics will be discovered 
at the LHC. One of the most popular candidate theories is weak scale supersymmetry. 
The superpartners of the SM particles, having different statistics, contribute to 
the radiative corrections to the Higgs mass with the opposite sign. 
In the limit of exact supersymmetry, all corrections to $m_h$ cancel. However, since 
supersymmetry must be broken, there is a remnant logarithmic divergence dominated by the 
negative contribution of the top quark. Then, if the soft supersymmetry breaking scale 
is around $1~$ TeV, the Higgs becomes tachyonic and the electroweak symmetry is 
radiatively broken. This scenario, in its simplest incarnation, the minimal supersymmetric
SM, is considerably constrained to live in a fraction of its parameter space. The
main constraint comes from the experimental lower bound on the Higgs mass. In the 
MSSM, $m_h$ is typically light, hardly above $120$~GeV. 

Another way to avoid the naturalness problem is not to have a Higgs at all, as in 
Technicolor (TC) theories. In these, a new strong interaction acting on techni-fermions,
makes them condense breaking the electroweak symmetry. It runs into trouble when 
trying to generate fermion masses through extended TC interactions, also felt by the 
SM fermions. These interactions give rise to flavor changing neutral currents (FCNC).
Raising the ETC energy scale may avoid the FCNC experimental bounds, but it makes
very hard giving masses to heavier fermions, particularly the top quark. In order
to address this, Topcolor interactions were introduced, giving rise to Topcolor-assisted
TC models. In addition to the fact that the picture, initially very appealing due to its
simplicity, has gotten rather complicated, there are additional bounds on 
TC theories coming from electroweak precision constraints (EWPC) which 
tell us that QCD-like TC theories are practically ruled out. 

If the Higgs boson is a composite state, and the compositeness scale is around the 
TeV, then the corrections to its mass are cutoff at the TeV scale. 
Models of this kind present several problems, particularly with EWPC. 
However, if the Higgs is a pseudo-Nambu-Godstone boson, 
its mass is protected by some global symmetry. The gauge interactions explicitly break
this symmetry giving rise to a Higgs mass.
Little Higgs theories, proposed recently, are aimed at putting the cutoff of these 
strongly interacting theories, at a scale of tens of TeV, whereas arranging for a 
collective symmetry mechanism that still protects $m_h$ from getting large. 
There are many choices of the global symmetry. In all cases, the gauge symmetry has
to be extended beyond that of the SM, and new fermions must be introduced. 
Thus, these models have a very rich phenomenology at the LHC.

Finally, and also recently, the possibility of solving the hierarchy problem 
in theories with extra spatial dimensions has been considered. We will concentrate 
on these proposals in the rest of this presentation.

\section{Extra Dimensions and the Hierarchy Problem} 
\label{extra} 
Although theories with compact extra spatial dimensions have existed for quite some time, 
particularly in the context of string theory, it was only recently that extra dimensions 
have been invoked as a solution to the gauge hierarchy problem. Here we present three
different scenarios where this is achieved.

\subsection{Large Extra Dimensions}

\begin{figure}
\includegraphics[width=5cm,height=7cm]{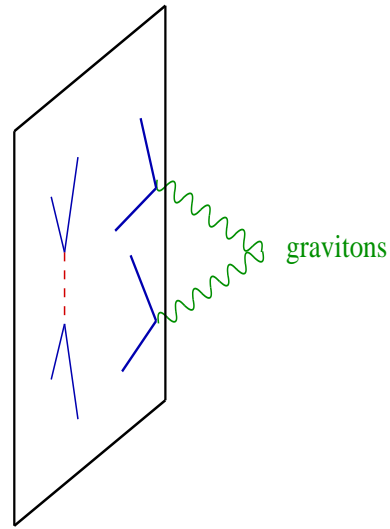}
\caption{Gravity propagating in the extra dimensional volume in the LED scenario. 
The SM matter and gauge fields are confined to a 4D slice, our brane. 
}
\label{brane}
\end{figure}

A new solution to the hierarchy problem involving extra spatial dimensions
was proposed in Ref.~\cite{led1}. In this setup, there are $3+n$ spatial dimensions, with 
$n$ compact dimensions of a typical compactification radius $R$. All matter and 
gauge fields are confined to a four-dimensional slice, a brane. On the other hand, gravity
can propagate in all $3+n$ dimensions, as shown in Figure~(\ref{brane}).
Therefore, 
gravity appears weak (i.e. $M_P\gg M_W$) 
because it propagates in the larger volume of compact extra dimensions, 
not available to the rest of the SM fields. The extra volume dilutes gravity's strength. 
The Planck scale is not a fundamental parameter in the extra-dimensional Einstein's action, 
but a scale derived from a volume suppression. The fundamental scale of gravity is $M_*$, 
and it satisfies 
\begin{equation}
M_P^2 = M_*^{n+2} \, R^n, 
\label{gauss}
\end{equation}
where $n$ is the number of extra dimensions, and $R$ is the average compactification radius. 
Then, in principle, the fundamental scale of gravity could be much smaller than $M_P$ 
if the extra dimensions are big enough. For instance, if we have in mind solving the hierarchy 
problem, then we could choose $M_*\simeq 1$~TeV. Then, 
\begin{equation}
R \sim 2\cdot 10^{-17}\,10^{\frac{32}{n}}{\rm cm}.
\label{rincm}
\end{equation}
Thus, if we take $n=1$, we get that $R\sim10^8~$Km, which is certainly excluded. 
Taking $n=2$, one has $R\sim 1$~mm, which is a distance scale 
already constrained by Cavendish-type experiments. For $n>2$, we need $R<10^{-6}$~mm, 
which is not going to be reached by gravity experiments any time soon.

\begin{figure}[h]
  \includegraphics[height=0.8in]{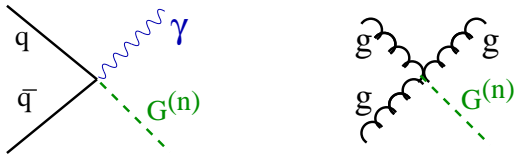}
  \caption{Production of KK Gravitons.}
\label{mono}
\end{figure}

The fact that gravity propagates in compact extra dimensions leads to the existence of 
graviton excitations with a mass gap given by $\Delta m\sim 1/R$. Then, in this scenario, 
there are new states, with spin 2, and with rather small masses. For instance, for $n=2$
the Kaluza-Klein graviton mass starts at about $10^{-3}$~eV, and for $n=3$ at about 
$100$~eV. Although the couplings of graviton excitations to matter are 
gravitationally suppressed, these states are so copiously produced at high energies 
($E\gg 1/R$) that when we sum over all these final states, the inclusive cross
sections are not suppressed by $M_P$, but by $M_*$:
\begin{equation} 
\sigma \sim \frac{E^{n}}{M_*^{{n}+2}}. 
\end{equation}
On the other hand, since KK graviton lifetimes are still $M_P^2$ suppressed they would 
escape detection, leaving large missing energy signals as their mark. The processes that most 
uniquely would point to this physics at hadron colliders are of the mono-jet type, as depicted 
in Figure~\ref{mono}.
Limits from the Tevatron already are available~\cite{karagoz} and are already around 
$O(1)$~TeV, depending on the number of dimensions. 
Another signal of LED comes from the virtual exchange of KK gravitons. 
This induces dim-6 operators of the form 
$(\bar q\gamma_\mu\gamma_5 q)(\bar f\gamma_\mu\gamma_5 f)$ 
entering, for example, in Drell-Yan production. 
Moreover, dim-8 operators ($T_{\mu\nu} T^{\mu\nu}$), result in 
$\bar ff\to\gamma\gamma, ZZ,\cdots$.
For $n=2$, the bounds on $M_*$ from the contributions of these operators are in the multi-TeV
region already~\cite{karagoz}. 

Astrophysical constraints have played an important role in the viability of the LED scenario.
The most tight bound comes from Supernova cooling, where graviton KK emission could cool 
the supernova too fast. For instance, for $n=2$ this requires $M_*>(10-100)$~TeV.

Finally, the proposal that the fundamental scale  of gravity, $M_*$, might not be far above
the TeV scale, raises the possibility that strong gravity effects -such as black holes- 
might be visible at colliders experiments~\cite{bhled}, 
or in ultra high energy cosmic rays~\cite{crled}.

\subsection{Universal Extra Dimensions} 

If  - unlike in the LED scenario - we allow fields other than the graviton to propagate in the 
extra dimensions, then constraints on $1/R$ are much more severe. This is because the couplings
of all other fields are not suppressed gravitationally, but at most by the weak scale. 
Naively, we would expect then that bounds on $1/R$ climb rapidly to $O(1)$~TeV, and this is what 
happens~\cite{karagoz} if {\em some} of the SM fields other than the graviton are in the bulk. 

\begin{figure}[h]
  \includegraphics[height=2cm]{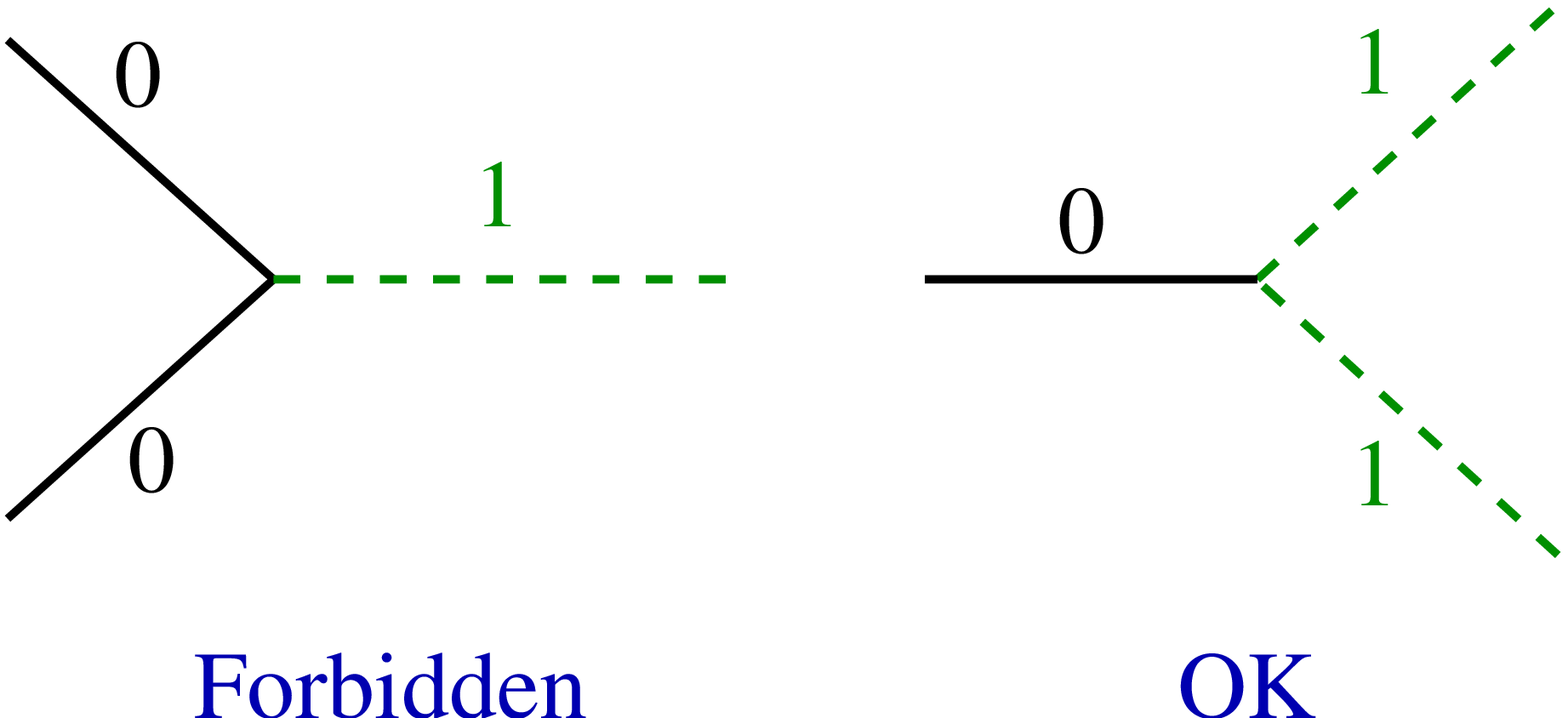}
  \caption{Orbifold Compactification.}
\label{kkncons}
\end{figure}

\begin{figure}[h]
  \includegraphics[height=0.8in]{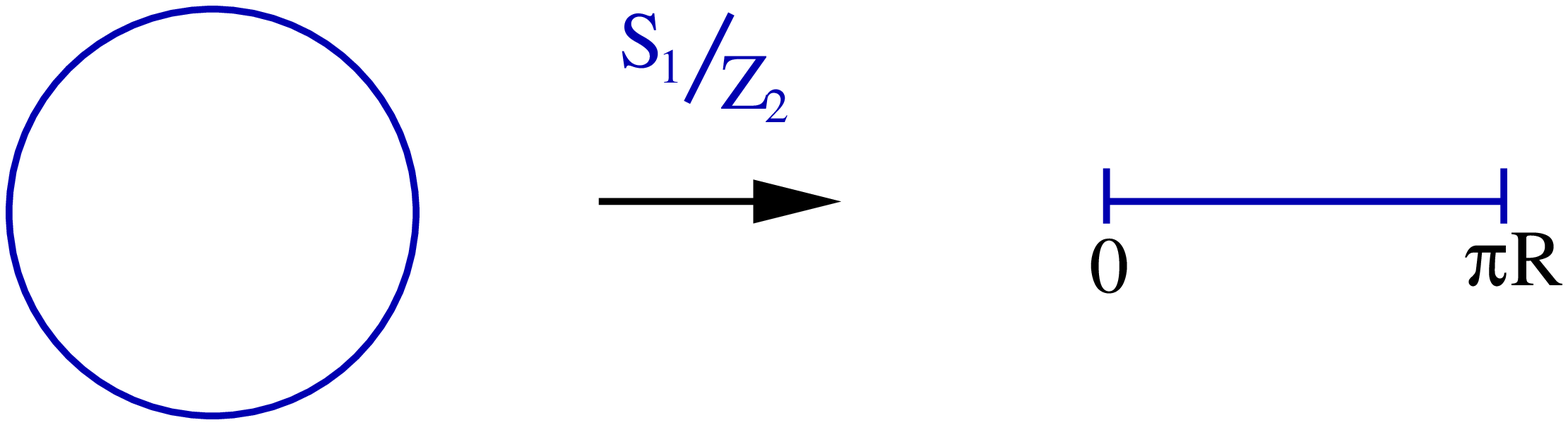}
  \caption{Orbifold Compactification.}
\label{orbi}
\end{figure}

However, it was shown in Ref.~\cite{ued1} that if {\em all } fields propagate in the extra 
dimensional bulk (universal extra dimensions), then bounds on $1/R$ drop to 
considerably lower values. The reason is that -upon compactification- momentum conservation leads
to KK-number conservation. For instance, in 5D, $p_5$ the fifth component of the 5D momentum
is quantified and given by 
\begin{equation}
p_5=n/R~,
\label{p5}
\end{equation}
with $n$ the KK number. The number $n$, (the fifth component of the momentum) must be 
conserved in interaction in the bulk. Thus, interactions involving two zero modes
and one of the first excitations or $1$ KK mode, are forbidden.
On the other hand, one zero mode can interact with two 1st excited states. 
This is illustrated in Figure~(\ref{kkncons}).
Among other things, this implies 
that in UED, KK modes cannot be singly produced, but they must be pair produced. 
This raises the bounds both from direct searches, as well as those from electroweak precision 
constraints~\cite{ued1}. 
Furthermore, compactification must be realized on an orbifold in 
order to allow for chiral fermion zero-modes, such as the ones we observe in the SM.
In 5D, for example, this means an $S_1/Z_2$ compactification as illustrated in Figure~\ref{orbi}.
As a consequence of the orbifolding, KK-number conservation is broken, but there is a 
remnant left:  the parity of the KK modes must be conserved at the vertexes. This 
KK-parity is very similar to R-parity in supersymmetric theories. It still means that 
zero-modes cannot be combined to produce a single KK excitation (for instance in an s-channel). 
In addition, KK-parity conservation means that the lightest state of the 
first KK level cannot decay into zero-modes, making it stable and a candidate for dark matter. 

Direct constraints from the Tevatron in Run I as well as electroweak precision constraints on 
oblique parameters, give 
\begin{equation}
1/R \geq \left\{\begin{array}{cc}
300{\rm ~GeV}&{\rm for ~5D} \\
(400-600){\rm ~GeV}  &{\rm for ~6D}
\end{array}
\right.
\nonumber
\end{equation}
These rather loose bounds imply that in principle even the Tevatron in Run II still has a chance 
of seeing UED. However, the signals are in general subtle. The reason is that -at leading order- 
all states in the same KK level are degenerate. Radiative corrections generate mass 
splittings~\cite{uedrad}, but these are still small enough for the energy release to be small
in the production and subsequent decay of KK states. 
For instance, if a pair of level 1 quarks is produced, each of them might decay as 
$Q_{1L}\to W^{\pm}_1Q'_L$, where the typical splitting between the $Q_{1L}$ and the $W^{\pm}_1$
might be just a few tens of GeV, depending on the number of extra dimensions (as well as the 
values of the brane kinetic terms). Thus, the quark jet tends to be soft, and this is repeated down 
the decay chain.  In Ref.~\cite{fooled}, the golden mode is identified as being 
$q\bar q\to Q_1Q_1\to Z_1Z_1 + \not E_T\to 4\ell + \not E_T$, where a large fraction of the 
missing energy is taken away by the lightest KK particle (LKP), in analogy with typical MSSM signals. 

In fact, the similarity with the MSSM makes distinguishing it from UED a challenging proposition. 
The most  distinct aspect of the UED scenario is the existence of KK levels, so the
observation of the second KK level might be what is needed to tell it apart from other 
new physics~\cite{fooled}. 
In the 5D case, for instance, level 2 KK states can decay via KK-number conserving interactions 
to either another lighter level 2 state 
plus a zero-mode, or to two level 1 states. In both cases, the energy release is very small. 
On the other hand, localized kinetic terms induce not only additional mass splitting, but also 
KK-number violating (but KK-parity conserving) interactions. These are volume suppressed, and 
therefore there will only be able to compete so much with the phase-space suppressed 
terms mentioned above. 

Six dimensional compactifications are also of interest. Unlike in the 5D case, in 6D the proton 
can be made naturally stable~\cite{uedsd}. Also in 6D the number of fermion families must 
be three. In 6D, however, there is more than one way of compactifying on an orbifold. 
We could have $T_2/Z_2$ or $T_2/Z_4$ compactifications. 
Recently, it was shown that an alternative way of folding the two compact extra dimensions, the 
chiral square, leads to the $T_2/Z_4$ KK theory~\cite{dp}. One can build gauge theories on the chiral 
square~\cite{bdp1} and then derive in it the spectrum and couplings of a given theory~\cite{bdp2}, 
in order to study its phenomenology. 

In the 6D case, the decay channel of the second KK level to two level 1 states is not 
present. This is because the generic mass of the second KK level is $M_2 = \sqrt{2}/R$, which is 
smaller than $2/R$, the sum of the masses of two level-1 states. 
Thus, in the 6D scenario, level 2 KK states 
can only decay through the localized kinetic terms and into two zero-modes.  
Furthermore, the signal also contains the contributions of the adjoint scalars, absent in 5D. 
These are the linear combinations of the 5th and 6th components of the gauge fields that are not
eaten by the KK modes. They decay almost exclusively to top pairs~\cite{bdp2}.
These signals then 
may be used to distinguish the 5D and 6D cases. 
Then, the 6D scenario, well-motivated in its own merits (e.g. proton stability~\cite{uedsd})
could be  distinguished form the more typical MSSM-like 5D case.

\subsection{Warped Extra Dimensions}

\begin{figure}[h]
  \includegraphics[height=1.8in]{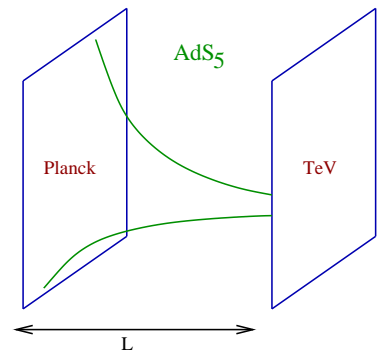}
  \caption{Warped Extra Dimension. The local cutoff is exponentially smaller than $M_P$.} 
\label{warped}
\end{figure}

A new solution to the hierarchy problem making use of one extra dimension was proposed in 
Ref.~\cite{wed1}. Unlike the two previous cases, the extra dimension does not have a flat metric. 
In the Randall-Sundrum (RS) setup, the metric is that of Anti-de Sitter in 5D, and is given by: 
\begin{equation}
ds^2 = e^{-2{k}|y|}\,\eta^{\mu\nu}dx_\mu dx_\nu  + dy^2,
\label{metric}
\end{equation}
which is a solution of Einstein's equations in 5D, as long as the bulk cosmological constant
is adjusted to cancel the cosmological constant on the fixed points. Then, the branes 
have a flat metric, as desired. In eqn.(\ref{metric}) $k\lta M_P$ is the $AdS_5$ curvature and 
$y$ is the coordinate of the fifth dimension. The only scale in the 5D 
Einstein-Hilbert action is $M_P$. However, when at a distance $y$ from the origin of the extra dimension,
all energies are exponentially suppressed by a factor of $\exp{(-ky)}$.
Then, if all SM fields, except gravity, were confined at a distance $L$ from the origin, the 
local cutoff would not be $M_P$ but 
\begin{equation}
\Lambda_L = M_P \,e^{-kL}.
\label{cutloc} 
\end{equation}
This is depicted in Figure~\ref{warped}. The compactification is done in the $S_1/Z_2$ orbifold, 
with $L=\pi R$. 
If we want the local cutoff to be the TeV scale, therefore explaining the hierarchy, we need to 
choose
\begin{equation}
k\,R\simeq O(10),
\label{kr10}
\end{equation}
which does not constitute a very significant fine-tuning. Then, in the RS scenario, an 
exponential hierarchy is generated by the non-trivial geometry of the extra dimension. 
This scenario already has important experimental consequences. Since gravity propagates 
in the bulk, there is a tower of KK gravitons. The zero-mode graviton has a wave-function 
localized towards the Planck brane and couples to matter with its usual coupling, suppressed by 
$1/M_P^2$. The KK gravitons, on the other hand, have masses starting at O(1)~TeV, and
couple to matter on the TeV brane as $1/(TeV)^2$. Then, KK gravitons can be produced at 
accelerators with significant cross sections. For instance, the Drell-Yan process 
would receive a contribution from s-channel KK gravitons as in 
$q\bar q\to G^{(n)}\to e^+e^-$.

The RS proposal solves the hierarchy problem because the radiative corrections to $m_h$ are
now cutoff at the TeV scale. The SM operates in our 4D slice. But the mechanism of EWSB is still
that of the SM. Moreover, the origin of fermion masses (the other hierarchy) is completely 
unexplained, together with a number of other issues ranging from gauge coupling unification to 
dark matter. Allowing additional fields to propagate in the bulk opens up a great deal of model 
building opportunities. In general, unless supersymmetry is invoked, the Higgs must 
remain localized on or around the TeV brane, or it would receive large quadratically divergent 
corrections of order of $M_P$, just as in the SM. 

If gauge fields are allowed in the bulk then their KK expansion takes the form
\begin{equation}
A_\mu(x,y) = \frac{1}{\sqrt{2\pi R}}\sum_{n=0}^{\infty} {A_\mu^{(n)}(x)} 
{\chi^{(n)}(y)}~,
\label{gaugekk}
\end{equation}
where $\chi^{(n)}(y)$ is the wave-function in the extra dimension for the 
nth KK excitation of the gauge field.  In the 4D effective theory, there is - in general - 
a zero-mode $A_\mu(x)^{(0)}$, and a KK tower of states with masses 
\begin{equation}
m_n \simeq (n -O(1))\times\pi{k} e^{-{k}\pi R},
\label{mn}
\end{equation}
starting at $O(1)$~TeV. Their wave-functions are localized towards the TeV brane. 
The gauge symmetry in the bulk must be enlarged with respect to the SM in 
order to contain an $SU(2)_L\times SU(2)_R$ symmetry. The extra $SU(2)_R$ restores 
a gauge version of custodial symmetry in the bulk, thus avoiding severe $T$ parameter 
constraints~\cite{adms} (in the dual language of the CFT, there is a global symmetry 
associated with it, the custodial symmetry). 

Just like the gauge fields, if fermions are allowed to propagate in the bulk they will
have a similar KK decomposition, given by
\begin{equation}
\Psi_{L,R}(x,y) = \frac{1}{\sqrt{2\pi R}}\,\sum_{n=0}\,{ \psi_n^{L,R}(x)} e^{2k\,|y|} 
{f_n^{L,R}(y)},
\label{fermions}
\end{equation}
where $f_n^{L,R}(y)$ are the wave-functions of the KK fermions in the extra dimension, and 
the superscripts $L$ and $R$ indicate the chirality of the KK fermion. Since fermions are 
not chiral in 5D, half of their components are projected out in the orbifold 
compactification. 
Unlike gauge fields in the bulk, fermions are allowed to have a mass term since  
there is no chiral symmetry protecting it. Then the typical, bulk fermion mass term looks like
\begin{equation}
S_f = \int d^4x~dy~ \sqrt{-g} \left\{ \cdots- ~{c}{~k} 
\bar\Psi(x,y)\Psi(x,y)\right\}~,
\end{equation}
where naturally $c\sim O(1)$, i.e. the bulk fermion mass is of the order of the 
AdS curvature scale $k$. The KK fermion wave-functions in this case have the form
\begin{equation}
{f_0^{R,L}(y)} = 
\sqrt{\frac{k\pi R\,(1\pm 2{c})}{e^{k\pi R(1\pm2{ c})}-1}}\;
e^{\pm{ c} \,k\,y}. 
\label{ferwf}
\end{equation}
Then, the localization of the KK fermion wave function in the extra dimension 
is controlled by the $O(1)$ parameter $c$ with {\em exponential} sensitivity~\cite{gp}. 
All that is needed to explain the wildly varying fermion spectrum is $O(1)$ flavor breaking 
in the bulk, which could be naturally originated at the cutoff. 
Fermions with wave-functions towards the TeV brane ($c<1/2$) will have a larger overlap with the 
Higgs VEV, and therefore a larger mass, of $O(v)$. Light fermions, on the other hand, will have 
wave-functions localized towards the Planck brane ($c>1/2)$. For $c=1/2$ the fermion 
wave-function is flat.
\begin{figure}[h]
  \includegraphics[height=1.8in]{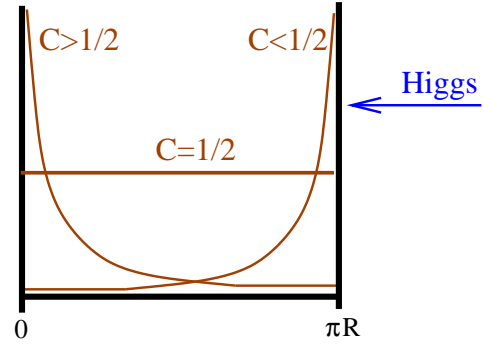}
   \caption{Fermions are localized according to the choice of the $O(1)$ parameter 
$c$, the bulk fermion mass in units of $k$, the AdS curvature.
}
\label{ferloc}
\end{figure}
This is shown in Figure~\ref{ferloc}. The need to generate a large enough value for $m_t$, 
forces us to localize the top quark not too far from the TeV brane. Even if we localize 
$t_R$ to this IR fixed point, the localization of $t_L$ -and consequently $b_L$- 
cannot be chosen to be at the Planck brane. If $b_L$ is forced to have a significant overlap 
with the TeV brane, there will be non-universal couplings of both the KK gauge bosons of all 
gauge fields, and the zero-mode weak gauge bosons ($W^\pm,Z$). 
The latter, result from the deformation of their (otherwise flat) 
wave-function  due to the Higgs VEV on the TeV brane. Fermions with profiles close to 
the TeV brane can feel this effect and couplings like $Z\to b_L\bar b_L$ 
would be affected~\cite{adms}. It would also lead to FCNCs at tree level mediated
by the $Z$, and that could be observed in $B$ decays such as $b\to\ell^+\ell^-$~\cite{bn}. 
On the other hand, the KK excitations of all gauge bosons would have non-universal couplings
to all fermions, and particularly to the top and to $b_L$. This could lead to interesting effects
in hadronic $B$ decays and $CP$ violation, especially when considering the interactions with 
KK gluons~\cite{gb}.

Finally, we could ask the question: could we get rid of the Higgs boson altogether ?
After all, it looks a bit ad hoc, localized in the TeV brane. We know that boundary conditions
(BC) can be used to break gauge symmetries in extra dimensional theories. 
It was proposed in Ref.~\cite{hless1} that the electroweak symmetry could be ``broken'' by 
BC in a 5D theory as a way to replace the Higgs field. The first question would be: what about 
the unitarity of electroweak amplitudes such as $W^+W^-$ scattering ? If the Higgs boson is not
present how are these amplitudes unitarized ? The answer is that the KK excitations of the 
gauge bosons do the job~\cite{cdh}. The actual models that (nearly) work  are similar 
to the one we had before on $AdS_5$, but without a Higgs boson on the TeV brane~\cite{hlads}. 
The BC can be thought of as obtained by the presence of brane-localized scalar fields that 
get VEVs. In the limit of these VEVs $\to\infty$, one recovers the BC. Thus although the 
origin of the BC might be a set of scalar fields getting VEVs, these need not be at low energies.
It is in this sense that the theory is Higgsless. 

\begin{figure}[h]
  \includegraphics[height=1.8in]{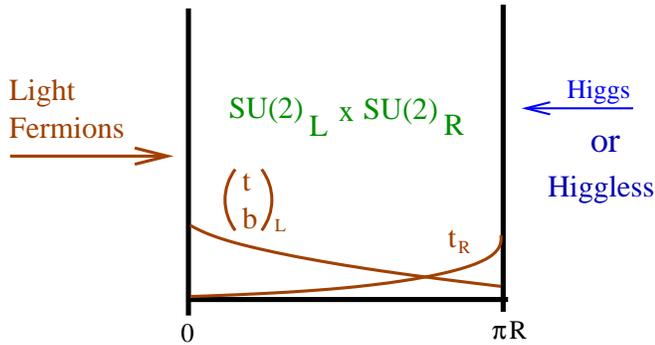}
   \caption{A Higgless model of EWSB and fermion masses. 
}
\label{newmods}
\end{figure}

The other question is how do fermions get their masses ? With the appropriate choice of 
BC the bulk gauge symmetry breaks as $SU(2)_L\times SU(2)_R\times U(1)_X \to 
U(1)_{\rm EM}$. But the BC restrict the gauge symmetry differently at different fixed points.
This can be seen in Figure~\ref{newmods}. The BC restrict the symmetry at the TeV brane to be
$SU(2)_{L+R}=SU(2)_V$, which allows us to write brane localized vector 
mass terms such as, for instance
\begin{equation}
{\cal S}_f = \int d^4x\int dy \delta(y-\pi R)\sqrt{-g}\,\left\{{M_u}q\psi_{\bar u} + 
{M_d}q\psi_{\bar d} + \cdots\right\},
\label{vecmass}
\end{equation}
where $M_u$ and $M_d$ are $\simeq O(1)$~TeV, and $q$, $\psi_{\bar u}$ and $\psi_{\bar d}$ a left-handed, 
right-handed up and right-handed down quarks respectively. Thus, the fermion mass hierarchy is 
still generated by $O(1)$ flavor breaking in the bulk fermion mass parameter $c$. The difference now is that
the masses are generated by the overlap with a vector mass term as opposed to the Higgs VEV. 
Thus from Figure~\ref{newmods} we see that the problem of flavor violation coming from the need to have 
a heavy top quark is still present here~\cite{bn}. 

\begin{center}
\begin{figure}[ht]
  \includegraphics[width=5cm]{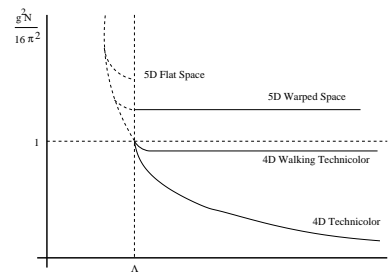}
   \caption{The 't Hooft coupling vs. energy scale for various theories. From Ref.~\cite{bn}.  
}
\label{phase}
\end{figure} 
\end{center}

The electroweak constraints on these kinds of theories are quite important. For instance, the $S$ 
parameter is given by~\cite{bn,bpr}
\begin{equation}
{S} \sim  16\pi\frac{v^2}{m_{KK}^2} = \frac{{N}}{\pi},
\label{svsn}  
\end{equation}
where in the second equality, $N$ refers to the size of the gauge group in the dual 4D CFT. 
Thus, for large $N$, which in the $AdS_5$ side corresponds to a weakly couple KK sector, the $S$
parameter tends to be larger than experimentally acceptable. 
Several possibilities have been considered in the literature to deal with this problem. For instance, 
negative $S$ contributions might be induced by TeV localized kinetic terms~\cite{ccgt}, however 
there is always a constraining combination of $S$ and $T$~\cite{sek}.
More recently, Ref.~\cite{ccgt2} advocates peeling light fermions off the Planck brane in order to reduce
$S$. This amounts to shift the couplings of light fermions to the gauge bosons, reabsorbing in the process
some, or even all, of $S$. 
Finally, one might take the result in eqn.~(\ref{svsn}) as an indication that $N$ must be small. This 
pushes the theory into the realm of a strongly coupled KK sector. This is the scenario entertained in 
Ref.~\cite{bn}. The result are theories where the KK sector is not well defined since KK states are not
narrow, well spaced resonances. In this case, there is no gap between the TeV scale and the cutoff 
of the 5D $AdS_5$ theory where we could defined individual, weakly coupled states. We would expect one 
broad resonance encompassing all the KK states. Above the cutoff of a few TeV, stringy dynamics come 
into play. This scenario is quite reminiscent of a Walking Technicolor theory. This can be seen in the 
schematic phase diagram of Figure~\ref{phase}, where the 't Hooft coupling $g^2 N/16\pi^2$ is plotted
against energy.
Here, $\Lambda$ is the energy scale where the CFT group exhibits non-trivial IR dynamics. 
In the large $N$ limit, it is possible to calculate $S$ reliably in the 5D $AdS_5$ theory. 
This is not the case in the Technicolor and Walking Technicolor theories. 
However, as $N$ is taken to be smaller, reliability may be lost in the 5D theory too. 
In any case, large or small $N$, the electroweak corrections come mostly from $E\sim \Lambda$, where
all theories will give similar results. However, at higher energies the theories may be quite different, 
with the dual of the $AdS_5$, a conformal theory with a 't Hooft coupling above 1 all to way to high 
energies. The question remains whether or not our knowledge of these differences can be put to use
to improve our understanding of strongly coupled theories at the TeV scale. 
Perhaps, EWSB is a consequence of a 4D conformal theory and the study of 5D theories could 
help illuminate some of its technical aspects.
In this direction, it is interesting to see how much of QCD at low energies can be obtained 
by assuming a theory with the IR at the GeV scale, as opposed to TeV. So far the results are 
remarkably good~\cite{adsqcd} when it comes to general aspects of the theory, such as the 
spectrum of vector resonances 
and their couplings. But we still need to understand why does this work for QCD. As far as we know, 
QCD is not a conformal or quasi-conformal theory. Why is then well described by RS models in $AdS_5$. 
The aim is to understand how to build strongly coupled theories at the TeV scale, by making use of 
weakly coupled theories in the warped background. We may learn something about QCD in the process.

\section{Conclusions}
The LHC is about to start exploring the energies where we expect the SM to fail. 
Although in principle is possible to have a weak scale only completed with the SM
light Higgs, this is very unnatural. More natural completions of the electroweak theory at the TeV
scale require the cancellation of the quadratic divergences in the Higgs mass.

Theories with extra spatial dimensions provide alternatives to supersymmetry and strong dynamics
at the TeV scale. 
We reviewed the main three ways in which these can solve the gauge hierarchy problem. 
In the case of Large Extra Dimensions, the true fundamental scale of gravity is the TeV scale,
whereas the Planck scale is derived from a volume suppression. Experimental signatures at colliders
are mainly the observation of various missing $E_T$ channels corresponding to the collective graviton 
states. 

Theories with Universal Extra Dimensions can have compactification radii still somewhat smaller than
the TeV. In them the fundamental scale of quantum gravity might reside in the tens of TeV. 
They present a very rich, albeit model dependent, phenomenology.  

Finally, the Randall-Sundrum scenario, where the extra dimension has a curved geometry corresponding
to $AdS_5$, may also give rise to a theory of flavor if matter and gauge fields are allowed to propagate
in the bulk. Various signals, both for its KK spectrum, as well as for flavor violating interactions
related to the origin of the flavor hierarchy, would be observed at the LHC. 
These kinds of extra dimensional theories could be viewed as a way to build strongly coupled 
4D theories of the TeV. Understanding the relation between a weakly coupled theory in 
$AdS_5$ and a strongly coupled theory in 4D should be an important step towards these models. 
Perhaps, modeling QCD at low energies in the $AdS_5$ framework will provide us with some clues
on how to go about studying a strongly coupled spectrum at the TeV, in the event this is what 
is observed at the LHC.

\end{document}